# The COVID-19 Inflation Weighting in Israel[*]

Jonathan Benchimol,[1] Itamar Caspi,[2] and Yuval Levin[3]

June 2022

**Abstract**

Significant shifts in the composition of consumer spending as a result of the COVID-19 crisis can complicate the interpretation of official inflation data, which are calculated by the Central Bureau of Statistics (CBS) based on a fixed basket of goods. We focus on Israel as a country that experienced three lockdowns, additional restrictions that significantly changed consumer behavior, and a successful vaccination campaign that has led to the lifting of most of these restrictions. We use credit card spending data to construct a consumption basket of goods representing the composition of household consumption during the COVID-19 period. We use this synthetic COVID-19 basket to calculate the adjusted inflation rate that should prevail during the pandemic period. We find that the differences between COVID-19-adjusted and CBS (unadjusted) inflation measures are transitory. Only the contribution of certain goods and services, particularly housing and transportation, to inflation changed significantly, especially during the first and second lockdowns. Although lockdowns and restrictions in developed countries created a significant bias in inflation weighting, the inflation bias remained unexpectedly small and transitory during the COVID-19 period in Israel.

[*] This paper does not necessarily reflect the views of the Bank of Israel. We thank participants from the Bank of Israel research seminars for their useful comments. We also thank the editor, the associate editor, and the referees for their constructive and useful comments.
The replication files for this paper are available here: github.com/yuval360/Covid-19-Inflation-Israel.
[1] Bank of Israel, Jerusalem, Israel. Corresponding author. Email: jonathan.benchimol@boi.org.il
[2] Bank of Israel, Jerusalem, Israel.
[3] Bank of Israel, Jerusalem, Israel.

# 1. Introduction

In recent years, the robustness of standard consumer price indices in measuring changes in purchasing power has been questioned. Inflation tends to be underestimated because consumers switch to cheaper items over time, and standard indices do not capture the effect of such switching. This substitution bias implies that inflation should be lower when measured with changing baskets than with fixed baskets since substituting away from goods whose prices rise disproportionately lowers inflation.

Braun and Lein (2021) show that the size of the biases can vary with economic conditions. While the substitution bias is shown to be relatively small, neglected preference adjustment and product entry/exit change the annual inflation figure. The substitution bias is substantial after a shock to relative prices. Our paper contributes to the discussion on substitution bias in general, particularly during the COVID-19 pandemic.

According to Jaravel and O'Connell (2020), consumer spending patterns in fast-moving consumer goods changed more dramatically during the pandemic than usual. However, they find that the degree of substitution bias is not greater than in previous years. During a pandemic, people tend to stock up on necessities even when prices are rising relative to other items. For instance, consumers' shopping behavior may have changed because of social distancing and the higher costs of searching during lockdowns. They may obtain the same products from more expensive outlets. In the reopening phase, the gap closes immediately since pre-COVID-19 habits are barely flexible.

In normal times, the Central Bureau of Statsitics (CBS) calculation of the Israeli Consumer Price Index (CPI) is based on fixed weights that are updated every two years according to the average consumption basket derived from the Household Expenditure Survey.[4] The use of fixed weights that change every two years makes sense if we want to measure the price change of a fixed basket of goods, rather than temporary changes in the cost of the basket that represent adjustments in consumption patterns due to, for instance, changes in relative prices (Cavallo, 2020).

The COVID-19 pandemic led to immediate and severe lifestyle changes for many consumers, including significant changes in consumption habits (Chetty et al., 2020; Goolsbee and Syverson, 2020). The lockdowns, restrictions, and changes in the economic behavior of the Israeli public led to a decrease in the proportion of goods and services consumed outside the home and in the weight of nonessential goods (Figure 1).

---

[4] The last update prior to the pandemic was made in January 2019, and another update was made in January 2021. In the latter, a limited number of components were adjusted to account for the impact of the COVID-19 pandemic based on credit card spending data.

**Figure 1** Change in Consumer Expenditures in Israel

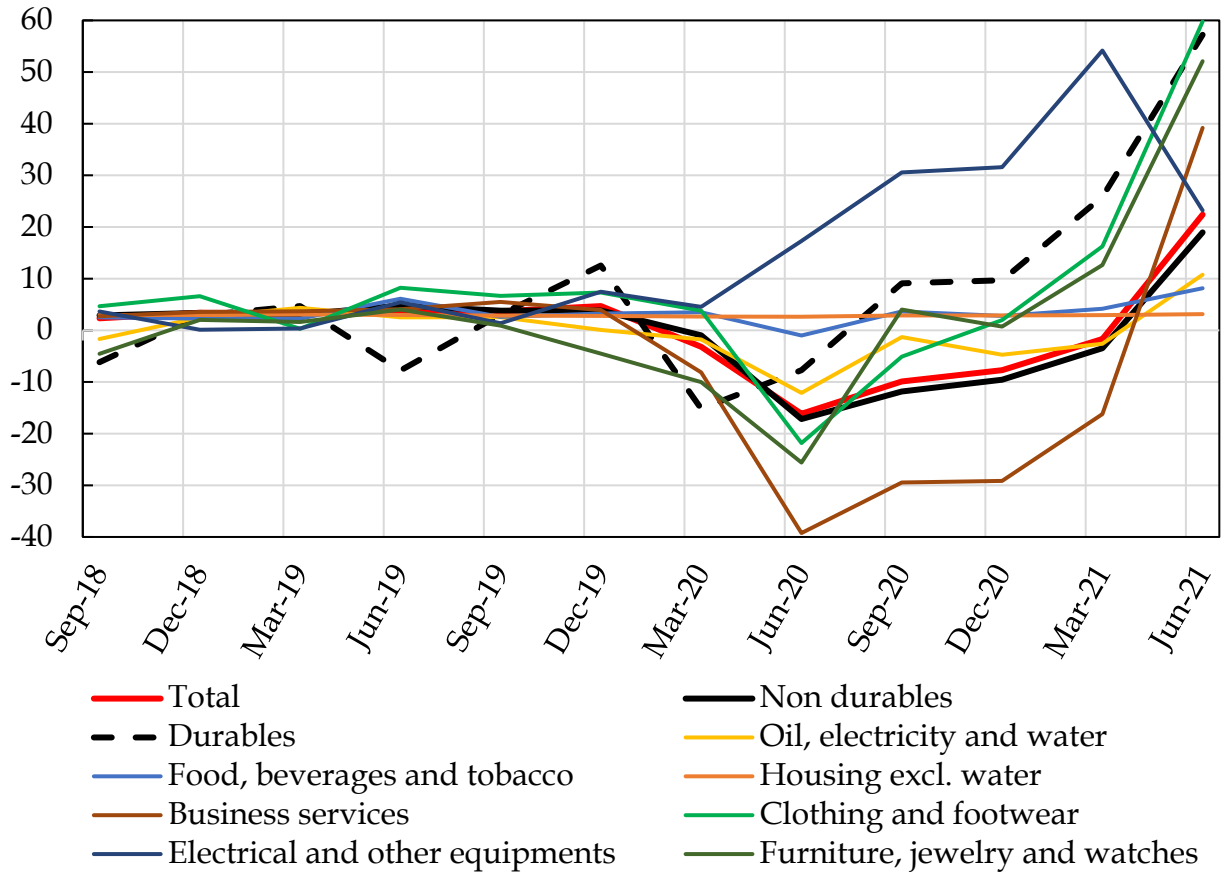

*Note*: The vertical axis represents the quarter-to-quarter percentage variation in private consumption and its components in Israel.
*Sources*: CBS and Bank of Israel.

According to Google Mobility Trends, there was a significant decline in movement to retail and recreation locations, which relates to lower demand for nonessential goods and services. Furthermore, a similar reduction in workplace presence corresponds to a reduction in the demand for transportation. Across all three lockdowns, the trend was similar, but the size differs. The first lockdown had the largest effect, and the last one had the smallest. The first lockdowns were characterized by tighter restrictions and greater enforcement, while the later lockdowns were characterized by fewer restrictions and less compliance (Bank of Israel Annual Report, 2021, pp. 21–31). This pattern is similar to that of other countries due to the nature of the pandemic and the response of governments (Carvalho et al., 2021). Thus, the weight of housing expenditures increased and the weight of transportation expenditures decreased within the average basket of goods and services, especially during the first and second lockdowns.[5] The vaccination campaign helped to lift restrictions and change consumption behavior again.

As a result of these changes, the CBS fixed consumption basket used in the calculation of the CPI in normal times does not represent the actual consumption basket during the COVID-19 period, and may therefore lead to a *biased weighting*, i.e., a biased calculation of inflation due to the significant shifts in the composition of the consumption basket during this period.[6]

By mid-January 2021, the number of COVID-19 cases and hospitalizations began to decline, with larger and earlier declines among the elderly. By early February, 45.3 percent (29.7 percent) of the total Israeli population and 89.9 percent (80 percent) of persons older than 60 years had received the first dose (both doses) of the vaccine or recovered from COVID-19 (Rossman et al., 2021). Although the Israeli vaccination campaign only started at the end of December 2020, by the end of March 2021 around 56 percent of Israelis, and more than 75 percent of those over 30, were vaccinated (both doses). Israel rapidly vaccinated its population, which allowed the government to ease restrictions quickly.

To measure the change in the composition of the household consumption basket and its effect on the measurement of inflation, we use data on credit card spending by industry to adjust the weights of the CPI and more accurately reflect consumption habits during the COVID-19 period.

Several recent studies have examined the biases in measuring inflation during the COVID-19 crisis. Our work is related to the literature on measuring weighting biases based on the use of dynamic weights estimated from consumption expenditure data. Cavallo (2020) uses credit- and debit-card spending data in the US to estimate the changes in the basket of goods during the COVID-19 crisis, and finds a negative bias in the official rate. In other words, COVID-19-adjusted inflation is significantly higher than the official rate. Assuming that the change in consumption in the US also occurred in other countries, Cavallo (2020) finds a negative bias in inflation for the month of May in most of the examined countries, although the size of the bias varies. Seiler (2020) presents additional support for these results using data on credit card spending in Switzerland. Based on a similar analysis, Reinsdorf (2020) strengthens the claim and compares data on changes in consumer behavior in Canada, groups countries geographically, and extends the sample period to February–May, which covers the entire first lockdown in most countries around the world. Overall, most of the bias in inflation estimates in most countries studied is due to increases in food spending and decreases in transportation spending.

---

[5] Spending on housing remained roughly constant, but its share of the basket grew while spending on the other components fell. By contrast, spending on transportation fell in absolute terms due to mobility constraints, but also relative to the other components.

[6] Additional biases in the CPI may arise from problems in measuring the CPI in the restricted industries and from changes in the quality of services consumed due to the associated health risk.

## 2. Methodology and Data

To estimate the change in the weight of an item in the consumption basket relative to the base months (January to February), or, in other words, to obtain the adjusted weight, we used the daily volume of credit card transactions.[7]

In particular, following Cavallo (2020), we multiply the monthly rate of change of each item (as measured by the CBS) by its COVID-19-adjusted weight to obtain its adjusted contribution to the inflation of the COVID-19 basket, as follows:

$$w^i_{COVID,t} = \frac{P^i_t Q^i_t}{\sum_i P^i_t Q^i_t} = \frac{w^i_{CBS} \Delta e^i}{\sum_i w^i_{CBS} \Delta e^i}, \qquad (1)$$

where $P^i_t$ and $Q^i_t$ are the prices and quantities of CPI category $i$ at time $t$, $w^i_{CBS}$ is the official weight of item $i$ in the calculation of the CPI, and $\Delta e^i = \frac{P^i_t Q^i_t}{P^i_0 Q^i_0}$ is the change in expenditures, where $P^i_0 Q^i_0$ and $P^i_t Q^i_t$ are the average credit card expenditure on item $i$ in the period January–February 2020 and in month $t$, respectively. Because the weights are relative, their importance in the basket can change even if their expenditure does not.

Since the classification of the credit card expenditures does not match the CBS classification of the CPI components[8], several modifications are required to transfer the changes in the weights of the credit card expenditures to the CPI weights. In addition to these modifications, we assume that expenditures on food (excluding meals eaten out) and fruits and vegetables change at the same rate, that expenditures on housing remain constant[9], and that expenditures on "other" goods and services change at the same rate as total expenditures.

---

[7] Our results do not account for changes in the volume of cash use during the crisis. For example, if the public did not change their spending on one of the items but switched to buying it in cash rather than with a credit card, there is a downward bias in the weight we calculate.

[8] Most of the main items in the CPI have been adjusted to match the parallel categories in the credit card data. The Housing Maintenance and Furniture and Household Appliances items were aggregated from two categories. The sub-item for meals eaten away from home was removed from the food category and added to culture and entertainment to improve compatibility with the credit card data.

[9] The CBS found that only a small proportion of leases (3 percent) had been reduced in April 2020. There has been no routine survey since then to determine whether rents have been reduced. However, frequent surveys conducted during lockdown periods that included a similar question found that only a small proportion of tenants had their rent reduced (0.3 percent).

## 3. Results

Figure 2 shows the share of CPI inflation weights by expenditure type and weighting scheme in 2020 and 2021. The weights, adjusted using the COVID-19 credit card expenditure data available in 2020 and 2021, vary more than the CBS weights between 2020 and 2021.

**Figure 2** Differences between CBS Official and COVID-19 Baskets

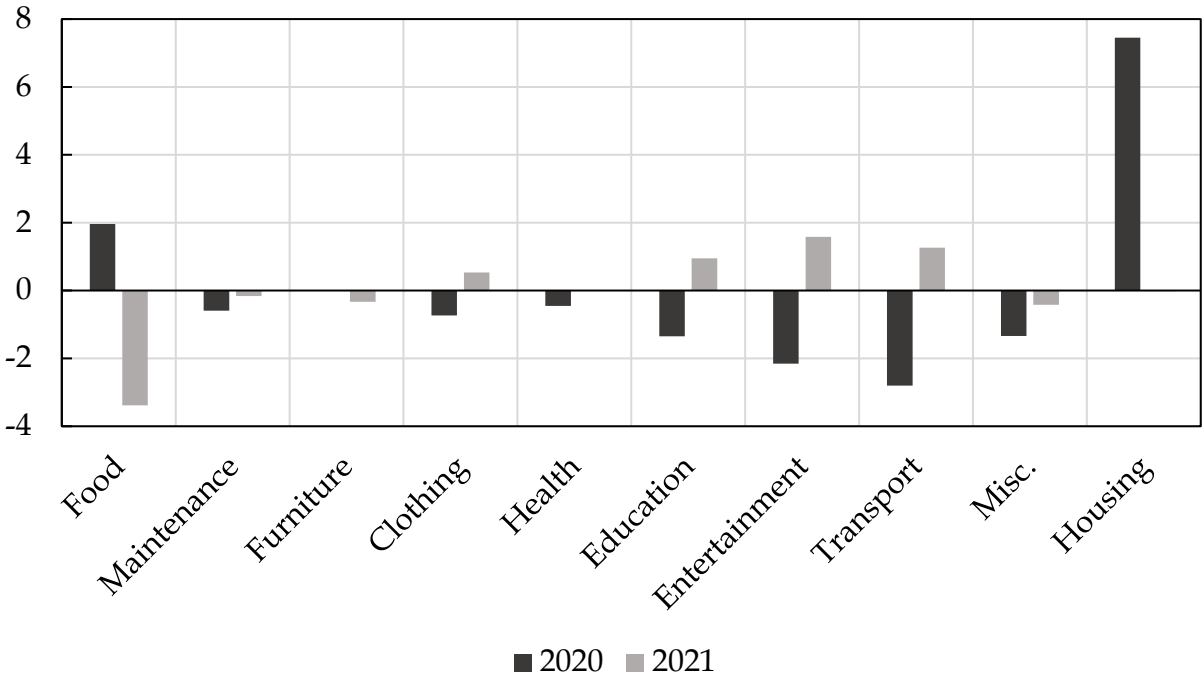

*Note*: The vertical axis represents the difference in percentage points of CPI inflation indices according to CBS and COVID-19 baskets calculated in January 2020 and 2021.
*Sources*: CBS and Bank of Israel.

Interestingly, the COVID basket weights decrease (increase) between 2020 and 2021 for housing (food and transportation) spending, while the CBS weights show the opposite movement. The disappearance of the spending spike in the housing spending share, along with the increase in spending shares in all other indices, that occurred in 2020 can be attributed to the return to regular consumption habits and thus to a normal basket of goods in the weeks following the so-far successful containment of the pandemic and the lifting of restrictions. The relatively high share of expenditure on food can be explained by the Passover holiday, which was celebrated at the end of March 2021.

Figure 3 shows the estimated evolution of the CPI weights. The influence and strength of the March–May 2020, September–October 2020, and January–February 2021 lockdowns are visible.

**Figure 3** Estimated Evolution of CPI Weights

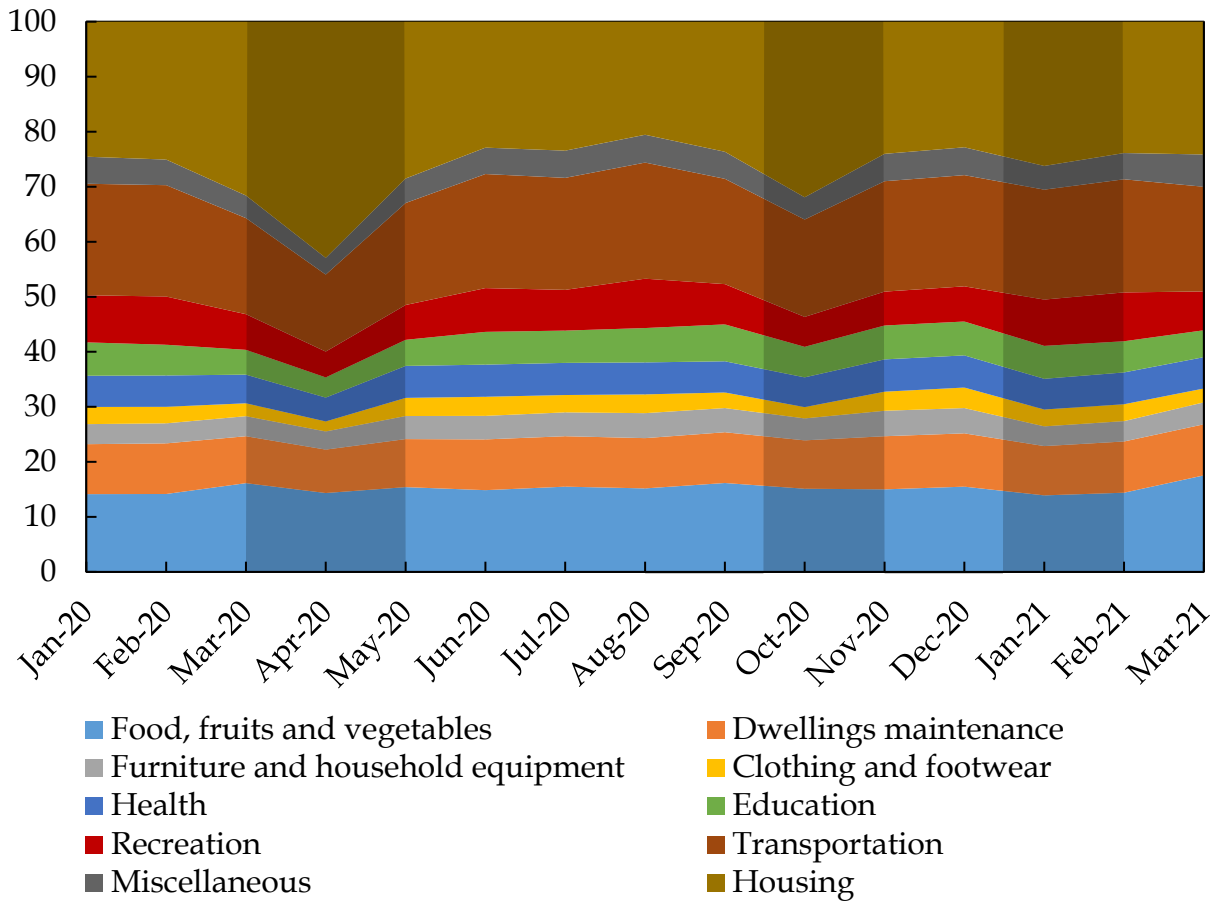

*Note*: The vertical axis represents percentage points. The shaded area presents the lockdown periods.
*Sources*: CBS and Bank of Israel.

However, there was no significant impact on the composition of the December basket. A decline in income and severe curbs on consumption caused spending on nonhousing basket goods to decline, resulting in a significant increase in the weight of housing expenditures. By contrast, the weights of the other items, especially transportation and entertainment, declined.

The upward trend in food and miscellaneous weights observed from December 2020 to March 2021 is mainly due to the Israeli vaccination campaign, which led to a gradual relaxation of restrictions.

Figure 4 shows the estimates of COVID-19 basket inflation alongside the official inflation rate on a monthly and annual basis.

**Figure 4** CPI Inflation under Official vs. COVID-19 Basket

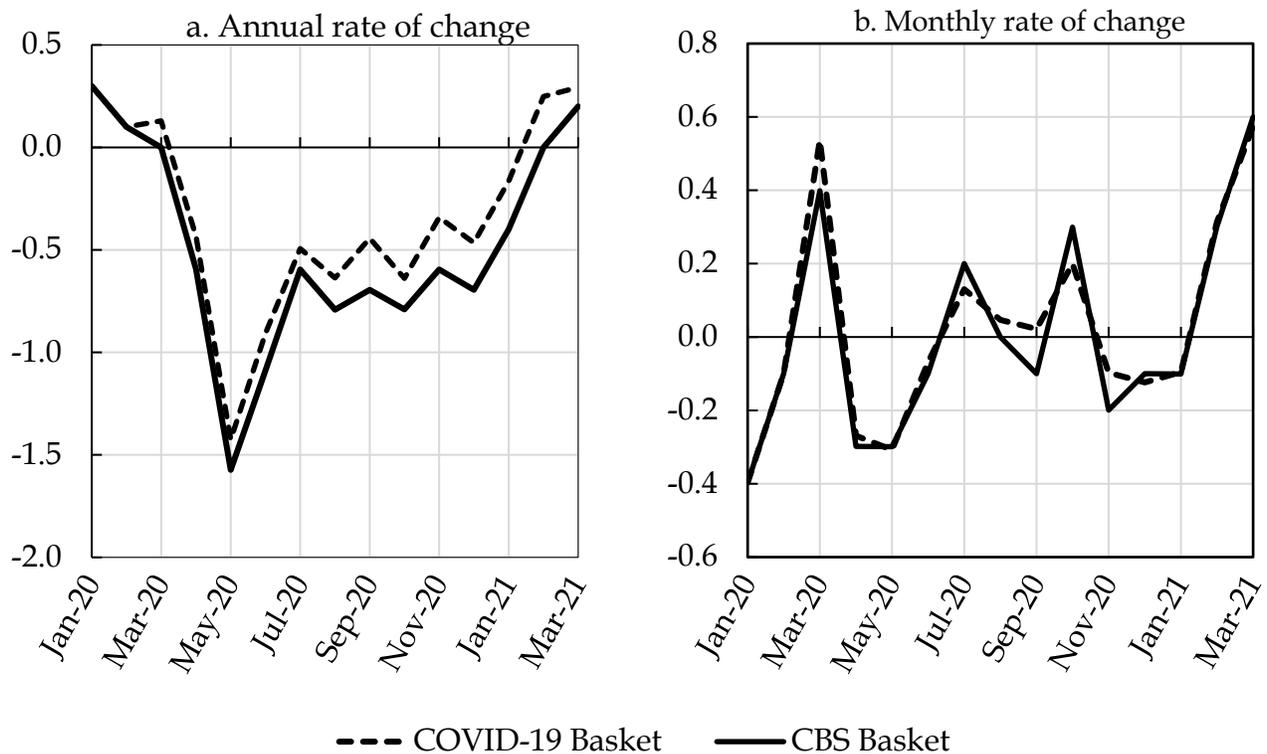

*Note*: The vertical axis represents percentage points.
*Sources*: CBS and Bank of Israel.

Figure 4b shows that the decline in transport use due to mobility restrictions, business closures, job losses, and remote working arrangements, together with the sharp decline in energy prices, led to a downward bias in inflation as measured by the CBS.[10] This downward bias was found in the official inflation rate on an annual basis. Although small in absolute terms, the finding may be theoretically important, as it shows that the weighting schemes can influence officially measured inflation. For example, the official yearly inflation rate is only about 0.2 percentage points lower than the COVID-19 basket inflation rate. In absolute terms, 0.2 percentage points may appear small, but it can be significant compared to overall inflation.

---

[10] Even when constant weights are assumed over the course of the entire year according to the consumption basket for April (the month in which there was the largest shock to the consumption basket), similar results are obtained, primarily with respect to the effect of the rise in energy prices that began in June.

By contrast, a look at the core index, which excludes food, fruits and vegetables, and energy[11]—items whose consumption was strongly affected relative to the other items—reveals a slight positive bias in the inflation measure, mainly due to the exclusion of energy. Figure 4a shows that the inflation bias persists until it starts to disappear in March 2021 due to the vaccination campaign that enabled a return to regular consumption habits.

The COVID-19 crisis and the accompanying economic restrictions were a global phenomenon, and the effects are likely to be similar in all countries. Figure 5 shows an international comparison of the weighting bias of inflation for the month of May 2020.

**Figure 5** International Comparison of the Inflation-Weighting Bias in May 2020

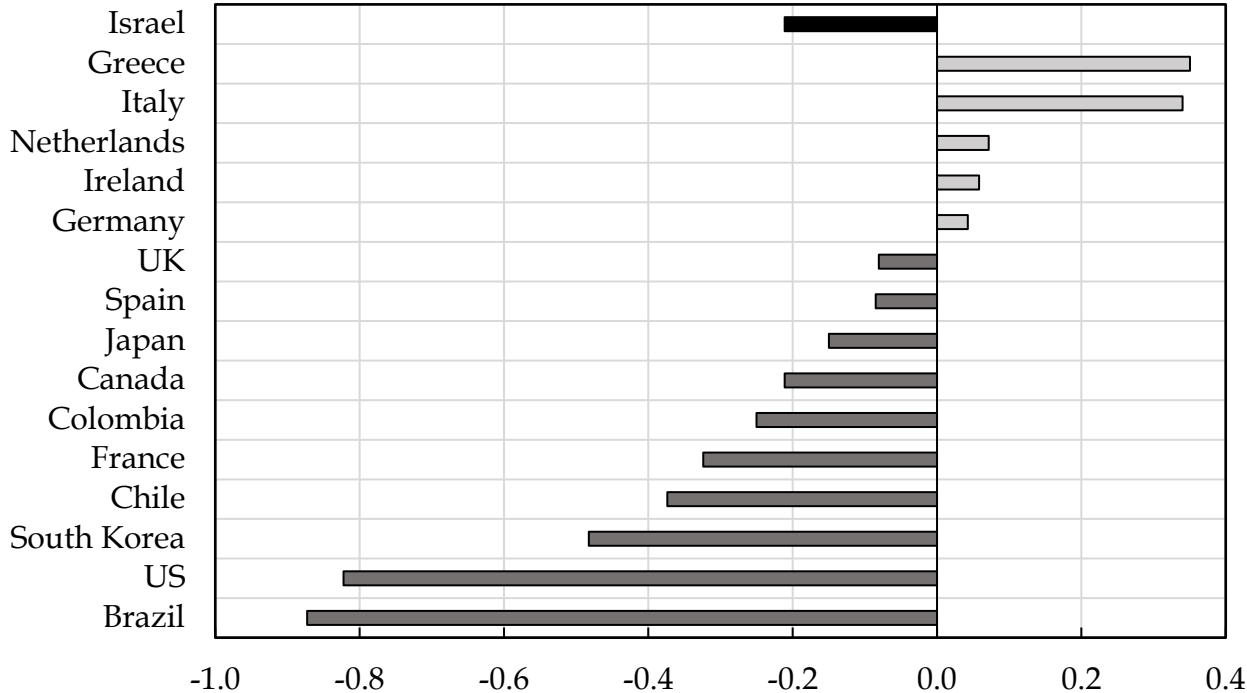

*Note*: The size of the bias represents the difference in percentage points between official and COVID-19 weights in May 2020. Gray (dark gray) bars represent a positive (negative) inflation bias. The inflation bias for Israel is in black.
*Sources*: Cavallo (2020) and Bank of Israel.

---

[11] Available upon request.

We calculated the figure for Israel, while the data for the other countries are taken from Cavallo (2020). One can see the heterogeneity in the size of the bias and even in its direction, although the bias is usually negative (i.e., the official inflation rate is lower than the COVID-19 basket inflation rate), which is also the case for Israel. This shows that the public and government response to the pandemic moderately affected the consumption basket.

## 4. Conclusion

As long as the pandemic continues, changes in the composition of the consumption basket in Israel and the rest of the world are likely to persist, especially during the peaks of the infection and restriction cycles. Therefore, they should be taken into account when considering short-term inflation. Nonetheless, we find that the successful Israeli vaccination campaign highlights the transitory nature of this bias, as it is helping bring back previous consumption habits. We also find that despite the shock to the consumption basket in Israel, which is similar in its trends to those generally observed in other countries, the bias in the overall inflation rate is relatively marginal.

**Bibliography**

Braun, R. and S.M. Lein (2021). "Sources of Bias in Inflation Rates and Implications for Inflation Dynamics." *Journal of Money, Credit and Banking*, 53(6), pp. 1553–1572.

Carvalho, V. M., J. Ramón García, S. Hansen, Á. Ortiz, T. Rodrigo, S. Rodriguez Mora, and P. Ruiz (2021). "Tracking the COVID-19 Crisis with High-Resolution Transaction Data." *Royal Society Open Science*, 8:210218.

Cavallo, A. (2020). "Inflation with Covid Consumption Baskets." NBER Working Paper #27352.

Chetty, R., J. N. Friedman, N. Hendren, and M. Stepner (2020). "The Economic Impacts of COVID-19: Evidence from a New Public Database Built Using Private Sector Data." NBER Working Paper #27431.

Goolsbee, A. and C. Syverson (2020). "Fear, Lockdown, and Diversion: Comparing Drivers of Pandemic Economic Decline 2020," *Journal of Public Economics*, 193:104311.

Jaravel, X. and M. O'Connell (2020). "High-Frequency Changes in Shopping Behaviours, Promotions and the Measurement of Inflation: Evidence from the Great Lockdown," *Fiscal Studies*, 41:3, pp. 733–755.

Reinsdorf, M. B. (2020). "COVID-19 and the CPI: Is Inflation Underestimated?" International Monetary Fund Working Paper #20/224.

Rossman, H., S. Shilo, T. Meir, M. Gorfine, U. Shalit, and E. Segal (2021). "COVID-19 Dynamics After a National Immunization Program in Israel." *Nature Medicine.*, 27, pp. 1055–1061.

Seiler, P. (2020). "Weighting Bias and Inflation in the Time of Covid-19: Evidence from Swiss Transaction Data," *Swiss Journal of Economics and Statistics* 156:13, pp. 1-11.